\documentclass[preprpint,5p]{elsarticle}

\usepackage{lineno,hyperref}
\modulolinenumbers[5]

\usepackage{units}
\usepackage{amsmath}

\newcommand{\pt}{p_\perp}
\newcommand{\jetpt}{p_\perp^\mathrm{(jet)}}
\newcommand{\jeteta}{|\eta^\mathrm{(jet)}|}
\newcommand{\jewel}{\textsc{Jewel}}

\journal{Physics Letters B}

\begin{document}

\begin{frontmatter}

\title{The role of initial state radiation in quenched jets}

\author{Korinna Zapp\fnref{ead}}
\fntext[ead]{korinna.zapp@thep.lu.se}
\address{Dept. of Astronomy and Theoretical Physics, Lund University,\\ Sölvegatan 14A, S-223 62 Lund, Sweden}
\address{Faculty of Science and Technology, University of Stavanger, Kristine Bonnevies vei 22, 4021 Stavanger, Norway}

\begin{abstract}
Jet quenching in heavy ion collisions and in particular the sub-structure of quenched jets are promising tools for investigating the microscopic processes  underlying jet quenching and the background medium's response to energy and momentum depositions. A quantitative understanding of the data can, however, be complicated by the presence of initial state radiation in reconstructed jets. Using an extended version of \jewel\ the effect of initial state radiation on different jet observables is studied in proton-proton and heavy ion collisions. It is shown that, depending on the observable and the jet radius, the initial state contributions can be sizable. Some general insights into when sizable effects can be expected also emerges.
\end{abstract}


\end{frontmatter}


\section{Introduction}

In heavy ion collisions jets are suppressed and have a modified internal structure compared to jets in proton-proton collisions due to energy loss of hard partons and induced radiation in the dense background (see~\cite{Casalderrey-Solana:2007knd,Wiedemann:2009sh,Majumder:2010qh,Mehtar-Tani:2013pia,Connors:2017ptx,Qin:2015srf,Cao:2020wlm,Cunqueiro:2021wls,Apolinario:2022vzg} for reviews).
Jet shapes and jet sub-structure observables are promising tools for gaining insights into the microscopic mechanisms of the interactions of hard partons and the background medium~\cite{Casalderrey-Solana:2011ule,Mehtar-Tani:2011hma,Blaizot:2014ula,Chien:2015hda,Casalderrey-Solana:2015bww,Dominguez:2019ges,Arnold:2020uzm,Mehtar-Tani:2021fud,Barata:2021byj,Andres:2022ndd}, as well as the medium's response to the energy and momentum deposited by hard partons~\cite{Cao:2020wlm,Neufeld:2011yh,Tachibana:2014lja,He:2015pra,Wang:2013cia,Cao:2016gvr,Casalderrey-Solana:2016jvj,KunnawalkamElayavalli:2017hxo}. A quantitative understanding of such observables requires a solid understanding of how the parton fragmentation process, (elastic and inelastic) re-scattering in the medium and medium response distribute the initial hard parton's energy and momentum in phase space. When jets are reconstructed using a jet finding algorithm they contain not only hard particles originating from a hard scattering process, but also other contributions that happen to be close to a fragmented hard parton. The latter consists of uncorrelated background from other nucleon-nucleon collisions and contributions originating from the same nucleon-nucleon interaction as the hard scattering, in particular initial state radiation and activity from multi-parton scattering. Initial state radiation (ISR) are emissions off partons entering the hard scattering and it is a consequence of the scale evolution of the parton distribution function (PDF). Its counterpart in the final state (FSR) are emissions off the hard scattered partons. In Monte Carlo event generators  both parts, ISR and FSR, are described by parton showers. It is important to note that initial state emissions can fragment by emitting final state radiation. The term ISR will here be used in the broader sense of including initial state emissions and their final state radiation.

ISR can give rise to jets and it can get clustered into jets composed predominantly of final state particles. The distinction between initial (IS) and final state (FS) jets, i.e. jets whose energy is carried mainly by ISR or FSR, respectively, is ambiguous, since any mixture of the two components is possible. For the qualitative considerations here it is assumed that there are two distinct jet populations: jets composed mainly of FSR with at most a modest ISR contribution, and vice versa. On general grounds one can expect that the contributions from ISR increase with jet rapidity, since ISR tends to be collinear with the beam. When ISR gets clustered into FS jets it forms a contribution that is only weakly correlated with the rest of the jet and has a wide distribution relative to the jet axis. It is therefore expected that the ISR contribution increases with jet radius (assuming anti-$k_\perp$ jets), because it scales with the jet area. These two properties, i.e.\ weak correlation with the jet and wide angular distribution, are also characteristic for medium response\footnote{Both ISR and medium response are not completely uncorrelated with the jet, because a higher jet energy will on average lead to both a somewhat higher level of ISR (a higher scale in the hard scattering opens up more phase space for ISR) and medium response.} and one may wonder whether ISR can fake signatures of medium response. Since ISR is semi-hard and jetty there is also a danger that it can be mistaken for medium induced radiation. 

It is the aim of this study to quantify the contributions from ISR in standard jet quenching measurements and get some qualitative insights to their importance. First results have been presented in~\cite{hilmisthesis}. This is in view of the fact that there is significant interest in reconstructing jets with large radii in heavy ion collisions~\cite{CMS:2021vui,ATLAS:2019rmd} in an endeavor to decode signs of induced radiation and medium response with increasing precision. If in a given observable the ISR contribution is not small, it should be included in the theoretical modeling (otherwise it would be difficult to draw conclusions from a comparison between theory and data). Parton shower based jet quenching models~\cite{ Armesto:2008qh,Schenke:2009gb,Casalderrey-Solana:2014bpa,Chen:2017zte,Putschke:2019yrg,Ke:2020clc} typically include ISR automatically, while in many analytical approaches it has to be added by hand.

\section{Event sample and analysis strategy}
\label{sec:analysis}

\begin{figure}
	\centering
	\includegraphics[width=7.5cm]{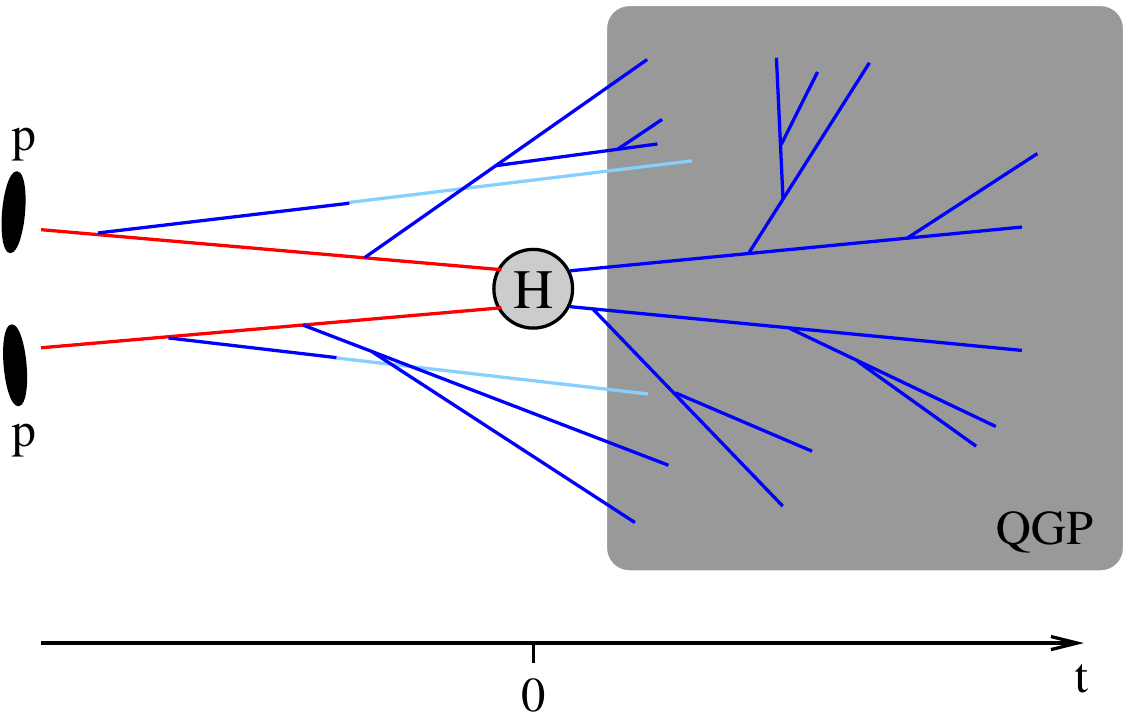}
	\caption{Schematic view of the parton showers in \jewel. Initial state partons with $Q^2 \le 0$ are shown in red and final state partons with $Q^2 \ge 0$ in blue (light blue lines are continuations of dark blue lines that are drawn in a different colour to indicate that they do actually not cross other lines). The hard scattering process depicted by the round gray blob happens at time $t=0$. A QGP (gray rectangle) forms shortly after the hard scattering and all partons entering it can interact with QGP constituents. The sketch is not to scale.}
	\label{fig:PSsketch}
\end{figure}

In previous versions of \jewel~\cite{Zapp:2012ak,Zapp:2013vla} ISR did not interact in the medium, i.e. it underwent vacuum evolution. Therefore, an extended version (\jewel-2.4.0\footnote{code available from \url{jewel.hepforge.org}}), in which ISR has the same medium modified evolution as FSR, is used here\footnote{The initial state parton shower, hard matrix elements and hadronisation are provided by \textsc{Pythia}\,6.4~\cite{Sjostrand:2006za} while \jewel\ simulates the final state parton shower and the interactions of the hard partons in the medium.}.
The parton showers and their time dependence are shown in a schematic way in fig.~\ref{fig:PSsketch}. The hard scattering has the highest scale in the event and takes place at time $t=0$. It is also the first process to be generated during event generation. After that the procedure in \jewel\ is to first run only the initial state parton shower. It starts at the hard scattering and evolves the incoming partons (red lines in fig.~\ref{fig:PSsketch}) backward in time and downward in scale $|Q^2|$ towards the incoming protons. In doing so the initial state parton shower emits partons off the incoming lines into the final state (blue lines in fig.~\ref{fig:PSsketch}). For emissions from the initial state parton shower the formation time is counted backwards from the hard scattering, i.e.\ they are emitted prior to the hard scattering. The final state partons radiated off initial state partons by the initial state parton shower have $Q^2 \ge 0$ and can emit further radiation. For final state radiation off initial state emissions the formation time is again counted forward. Since they start out at times earlier than the hard scattering, one or several final state emissions can take place in vacuum before the formation of the QGP medium. Once the partons enter the medium they interact in exactly the same way as final state emissions do. In the initial state parton shower the scale increases towards the hard scattering, which means that high $\pt$ partons are on average emitted closer to the hard scattering and therefore later than partons with low $\pt$. The harder partons thus have less time before the QGP forms and the amount of medium modifications one can expect is similar to those of hard parton emitted in the final state. Softer emissions from the initial state, on the other hand, form on average long before the QGP forms and experience much less modifications due to interactions in the medium.

\smallskip

Two di-jet samples at $\sqrt{s_\mathrm{NN}} = \unit[5.02]{TeV}$ were generated: one in \unit[0-10]{\%} central Pb+Pb collisions with the EPPS16NLO nuclear PDF set~\cite{Eskola:2016oht} and the corresponding p+p sample with CT14NLO~\cite{Dulat:2015mca} PDFs. Both PDF sets are provided by \textsc{Lhapdf}~\cite{Buckley:2014ana}. The Pb+Pb sample includes recoils, i.e.\ the effects of medium response. The subtraction of the corresponding thermal momenta is performed using the new constituent subtraction method~\cite{Milhano:2022kzx} and recoils from interactions of the ISR component are also labeled as ISR. The background is modeled with \textsc{Jewel}'s standard simplified background model ($T_\mathrm{i}=\unit[590]{MeV}$ and $\tau_\mathrm{i}=\unit[0.4]{fm}$~\cite{Shen:2014vra}.)

The events are analysed at parton level, since after hadronisation it is no longer possible to assign a particular hadron to either the initial or final state component of the event. The events are analysed using Rivet~\cite{Bierlich:2019rhm}. Jets are reconstructed using the anti-$k_\perp$ algorithm~\cite{Cacciari:2008gp} provided by the \textsc{FastJet} package~\cite{Cacciari:2011ma} with different radii. Unless stated otherwise, jets with $\jetpt > \unit[100]{GeV}$ and $\jeteta < 3$ are considered.

\section{Global impact of initial state radiation on jets}
\label{sec:res1}

\begin{figure}
	\includegraphics[width=8.5cm]{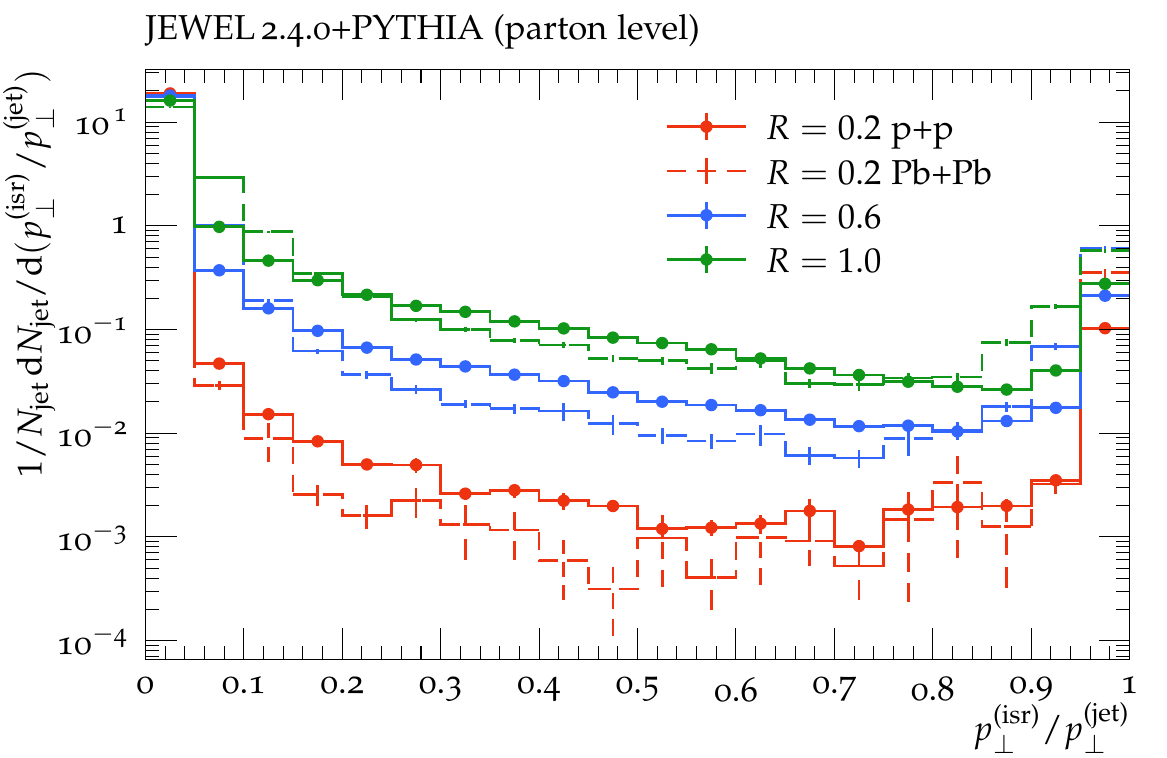}
	\caption{Fraction of the jet transverse momentum $\jetpt$ carried by ISR partons for different jet radii in p+p (solid histograms with markers) and central Pb+Pb collisions (dashed histograms).}
	\label{fig:ptfrac}
\end{figure}

Fig.~\ref{fig:ptfrac} shows the distribution of the fraction of the jet transverse momentum $\jetpt$ carried by ISR partons for different jet radii. There are two maxima at zero and one, corresponding to FS and IS jets with only a small admixture of other partons, respectively. Between them is a broad continuum of jets that are not predominantly composed of only one type of partons. The importance of this continuum naturally increases with jet radius. Jets with $p_\perp^\mathrm{(isr)}/\jetpt > 0.5$ are classified as IS jets and those with $p_\perp^\mathrm{(isr)}/\jetpt < 0.5$ as FS jets. Choosing a different value for the separation leads to quantitative changes in some distributions, but qualitatively the picture does not change. 

\begin{figure}
	\includegraphics[width=8.5cm]{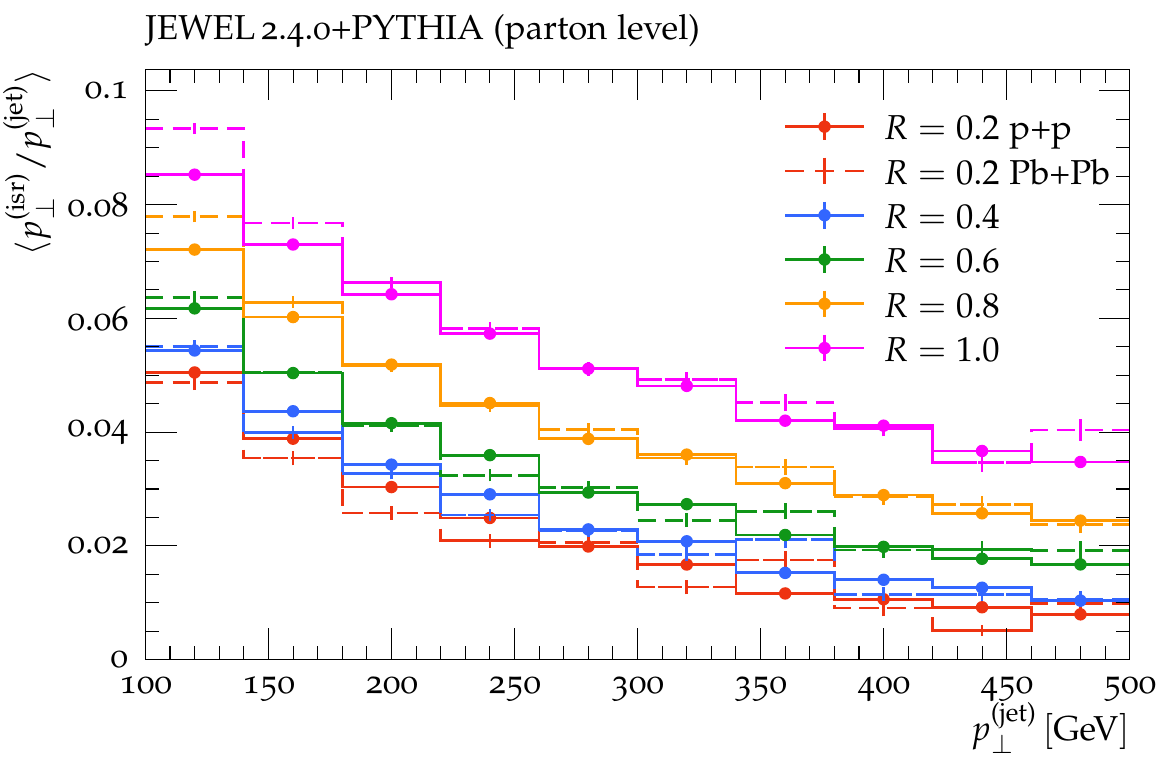}
	\caption{Average fraction of the jet transverse momentum $\jetpt$ carried by ISR partons versus jet $p_\perp$ for different jet radii in p+p (solid histograms with markers) and central Pb+Pb collisions (dashed histograms).}
	\label{fig:ptfrac_vs_pt}
\end{figure}

\begin{figure}
	\includegraphics[width=8.5cm]{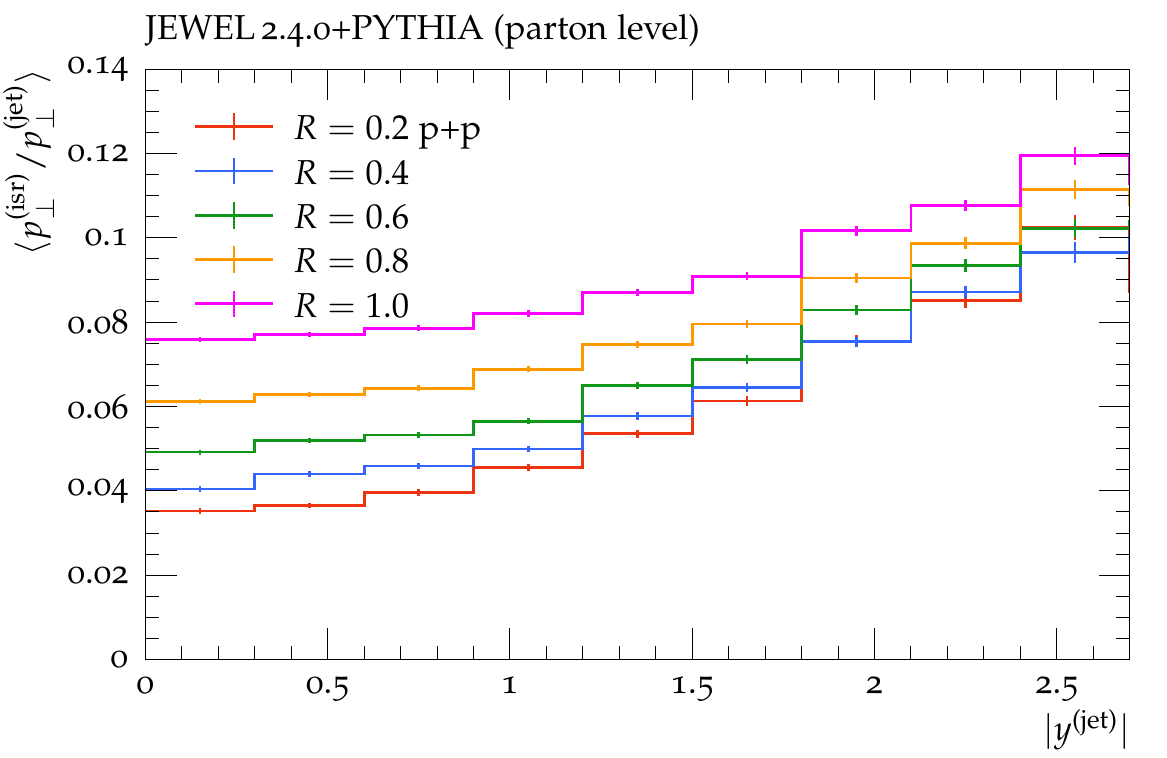}
	\caption{Average fraction of the jet transverse momentum $\jetpt$ carried by ISR partons versus jet rapidity for different jet radii in p+p collisions.}
	\label{fig:ptfrac_vs_y}
\end{figure}

The average ISR $p_\perp$ fraction as a function of $\jetpt$ (fig.~\ref{fig:ptfrac_vs_pt}) shows roughly the $1/\jetpt$ falloff expected if the ISR contribution is independent of the jet's $p_\perp$. As expected, the ISR contribution increases with jet radius. There are only small differences between vacuum and medium-modified jets. The dependence on rapidity (fig.~\ref{fig:ptfrac_vs_y}) is therefore shown only for p+p collisions. It shows the expected increase with rapidity. This increase does, however, not imply that the size of the ISR contribution depends strongly on the rapidity range considered, since the jet cross section drops quickly as a function of rapidity. Reducing the coverage to $|\eta^\mathrm{(jet)}|<2$, for instance, reduces the jet cross section by $\unit[10]{\%}$ and consequently changes the results presented below by at most a few percent.

\begin{figure}
	\includegraphics[width=8.5cm]{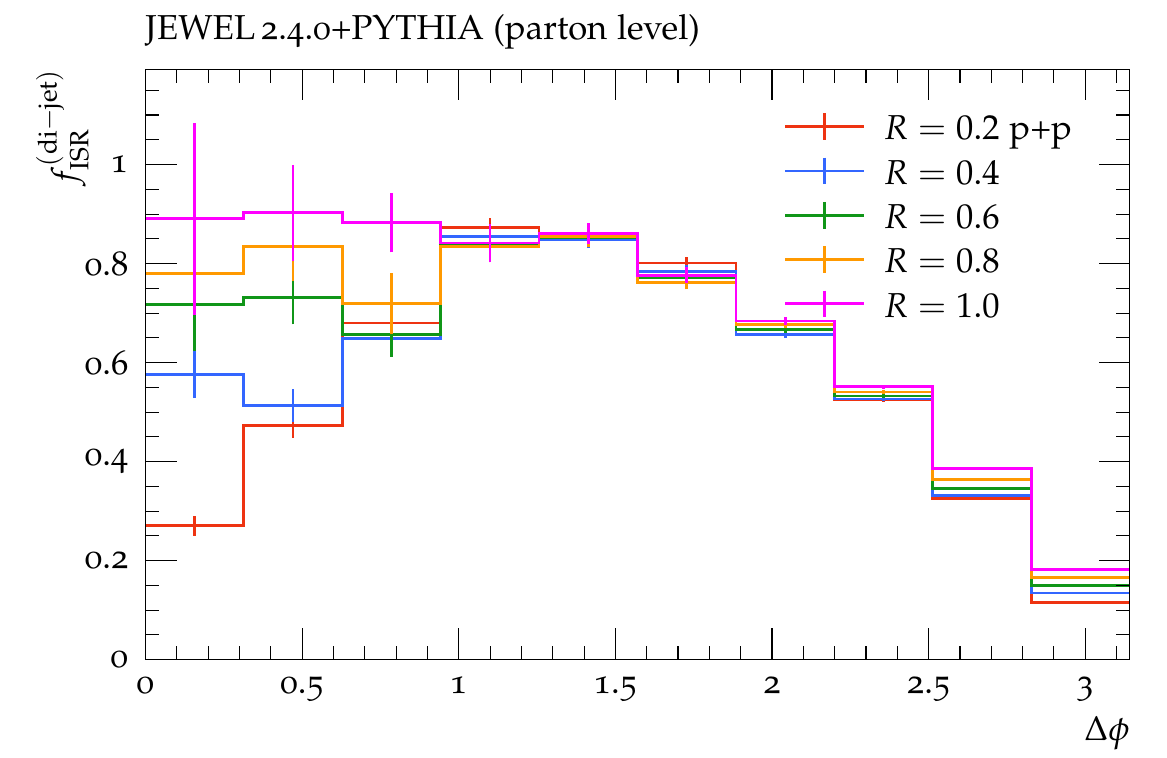}
	\caption{Fraction $f_\mathrm{ISR}^\mathrm{(di-jet)}$ of di-jets with at least one IS jet versus azimuthal angle $\Delta \phi$ between the jets.}
	\label{fig:isfrac_dijets}
\end{figure}

Overall, the ISR contribution to the jet $p_\perp$ is at the level of a few percent and reaching $\unit[10]{\%}$ at $|y^\mathrm{(jet)}| \approx 2.5$. The fraction of IS jets (defined as jets that receive more than half of their transverse momentum from ISR) in the sample is also at the level of a few percent.  Globally, the impact of ISR on single-inclusive jets is negligible. However, already for di-jets this is not necessarily the case, as can be seen in fig.~\ref{fig:isfrac_dijets}, which shows the fraction $f_\mathrm{ISR}^\mathrm{(di-jet)}$ of di-jets where at least one of the jets is an IS jet as a function of the azimuthal angle $\Delta \phi$ between the jets. It should be noted that for the di-jet selection the $p_\perp$ requirement for the sub-leading jet is relaxed to $\jetpt > \unit[60]{GeV}$. At $\Delta \phi \approx \pi$ the sample is dominated by FS jets, albeit with a sizable contribution of IS jets. Around $\Delta \phi \approx \pi/2$ mixed configurations with one IS and one FS jet make up $\unit[80]{\%}$ of all di-jets (the contributon from double IS configurations is negligible). At small angles the picture depends on the jet radius: for small radius jets it is more likely that the fragments of the same hard parton end up in two separate jets giving a FS-FS configuration. For large radius jets this option does not exist and the sample is dominated by configurations containing IS jets. Again, the results for Pb+Pb collisions are not shown because they are similar to p+p. Deflected jets have been argued to be signs for Moli\`ere scattering in the medium~\cite{DEramo:2018eoy}. The results of fig.~\ref{fig:isfrac_dijets} indicate that when looking for deflected jets ISR has to be taken into account.

\section{Impact of initial state radiation on jet shapes and jet sub-structure}
\label{sec:res2}

\begin{figure}
	\includegraphics[width=8.5cm]{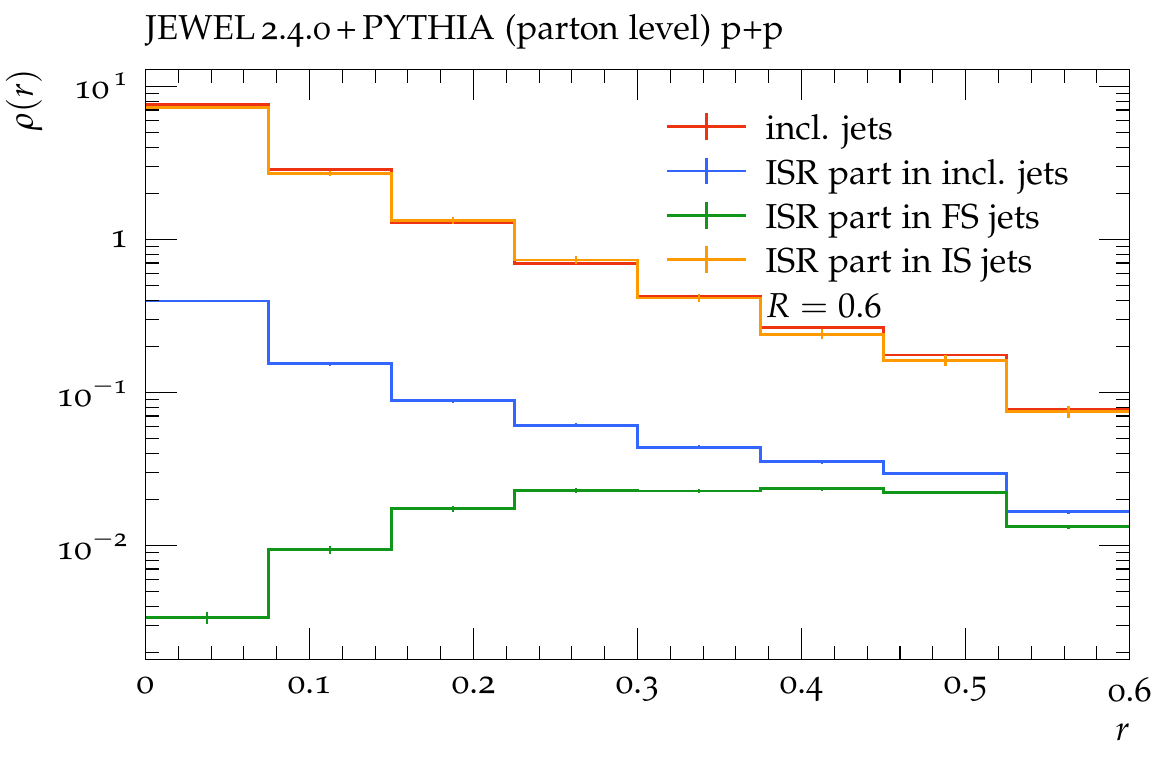}
	\caption{Jet profile for inclusive jets (red) and ISR contribution to the jet profile in inclusive jets (blue), FS jets (green) and IS jets (orange) in p+p collisions, $R=0.6$.}
	\label{fig:jetprofile_pp}
\end{figure}

The jet profile $\rho(r)$ is the fraction of the jet's $\pt$ contained in an annulus of size $\delta r$ located at a distance $r$ from the jet axis. It is defined here in a slightly different way as a jet-hadron correlation
\begin{equation}
	\rho(r) = \frac{1}{\jetpt} \hspace{-2mm} \sum_{\substack{k \mathrm{\ with\ }\\ \Delta R_{kJ} \in [r, r+\delta r]}} \hspace{-3mm} \pt^{(k)} \,,
\end{equation}
where the sum is over all particles in the event (and not only over the jet constituents), and $\Delta R_{kJ} = \sqrt{\Delta \phi_{kJ}^2 + \Delta y_{kJ}^2}$ is the angular separation between particle $k$ and the jet axis. For $r\le R$ the difference between the two definitions is minimal.

The red histogram in Fig.~\ref{fig:jetprofile_pp} shows the jet profile in p+p collisions for inclusive $R=0.6$ jets, while the blue histograms represents the ISR contribution to the jet profile. The latter is actually a mixture of two very different distributions, as the sample includes both FS and IS jets. The green and orange histograms show the ISR contribution to the jet profile in FS and IS jets separately. In FS jets the ISR contribution carries only a small fraction of the total jet $\pt$ and rises strongly with $r$ because the area of the annuli increases with $r$. In IS jets, on the other hand, the ISR part carries essentially all of the jet momentum and has the same shape as the inclusive distribution dictated by the perturbative jet evolution.

\begin{figure}
	\includegraphics[width=8.5cm]{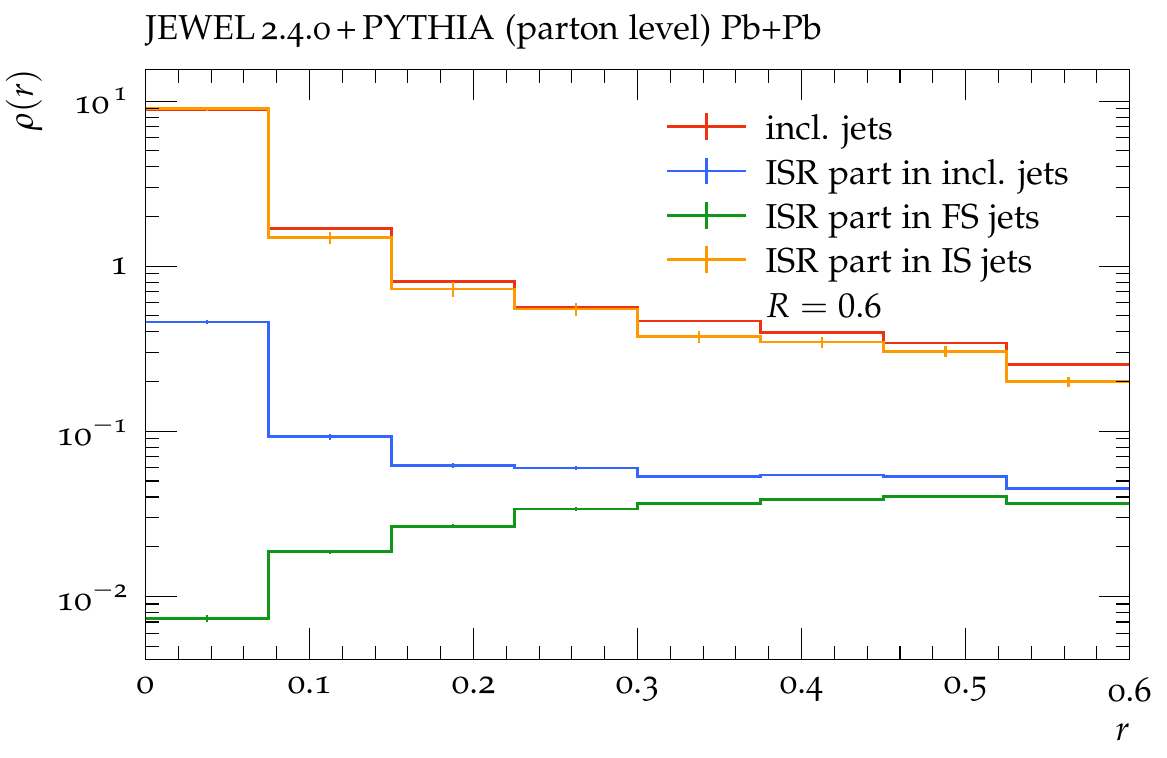}
	\caption{Jet profile for inclusive jets (red) and ISR contribution to the jet profile in inclusive jets (blue), FS jets (green) and IS jets (orange) in central Pb+Pb collisions, $R=0.6$.}
	\label{fig:jetprofile_PbPb}
\end{figure}

\begin{figure}
	\includegraphics[width=8.5cm]{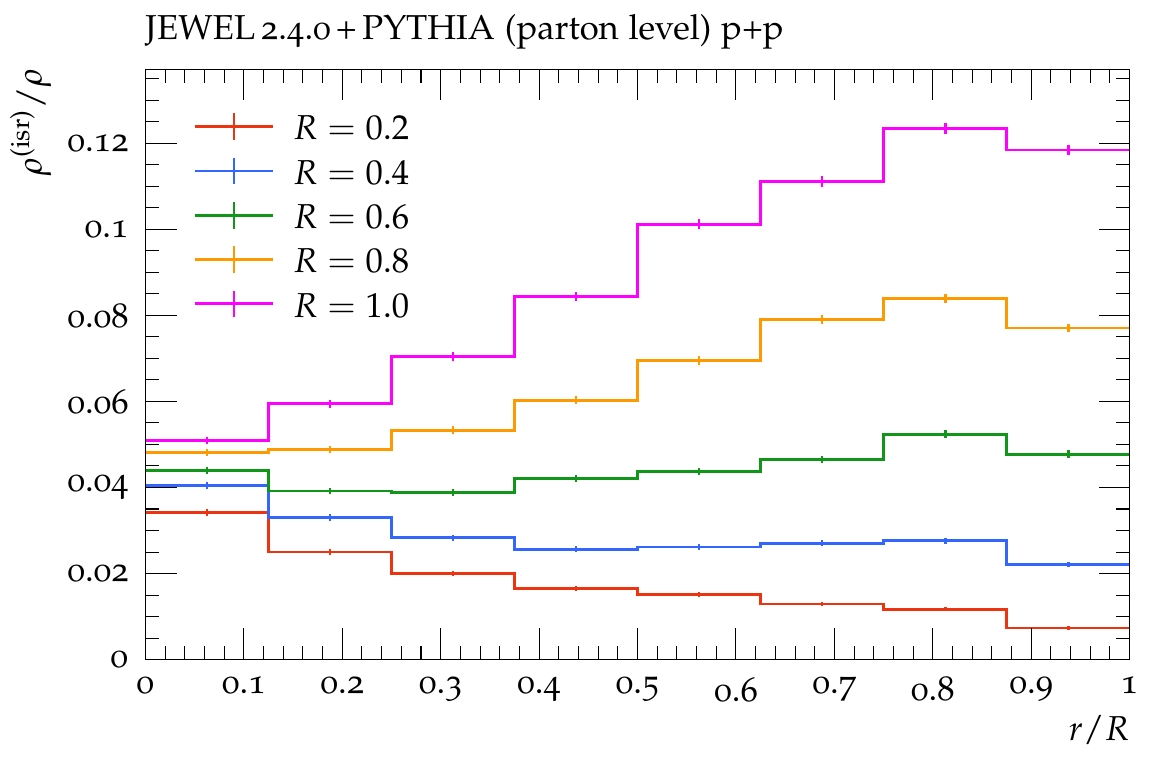}
	\caption{Ratio of ISR contribution to full jet profile in inclusive jets for different jet radii in p+p collisions.}
	\label{fig:isrjetprofile_pp}
\end{figure}

\begin{figure}
	\includegraphics[width=8.5cm]{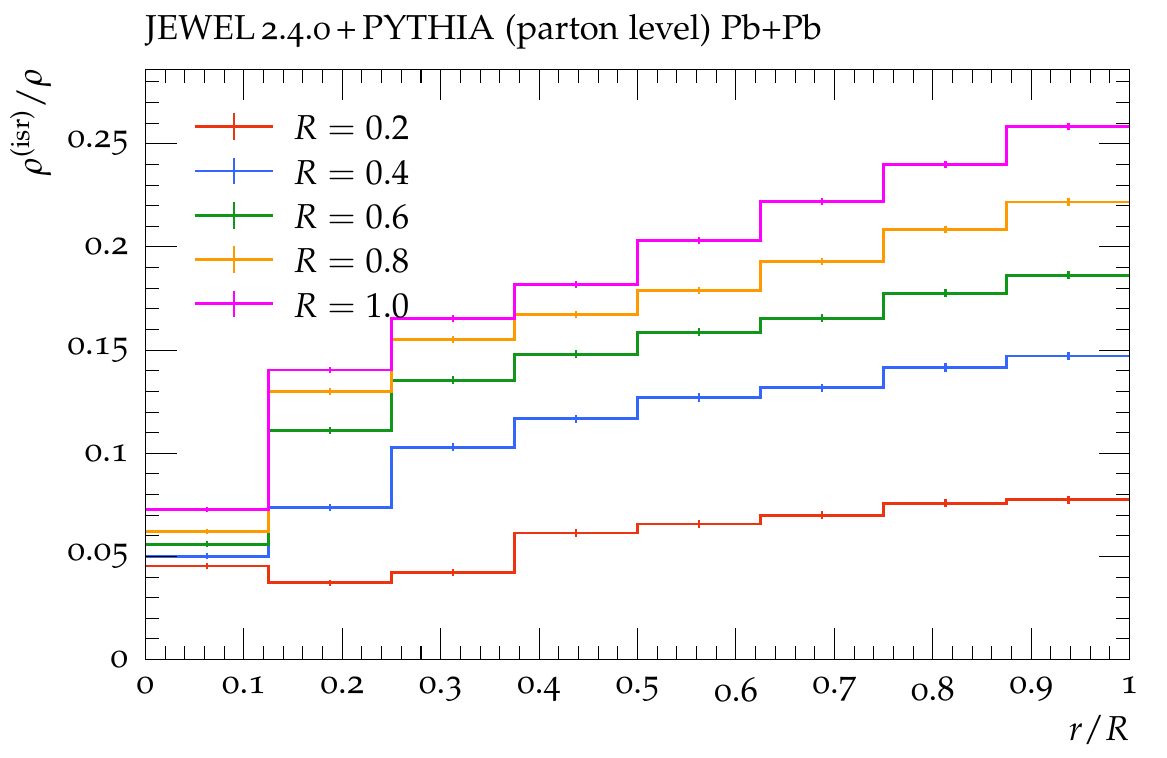}
	\caption{Ratio of ISR contribution to full jet profile in inclusive jets for different jet radii in central Pb+Pb collisions.}
	\label{fig:isrjetprofile_PbPb}
\end{figure}

\begin{figure}
	\includegraphics[width=8.5cm]{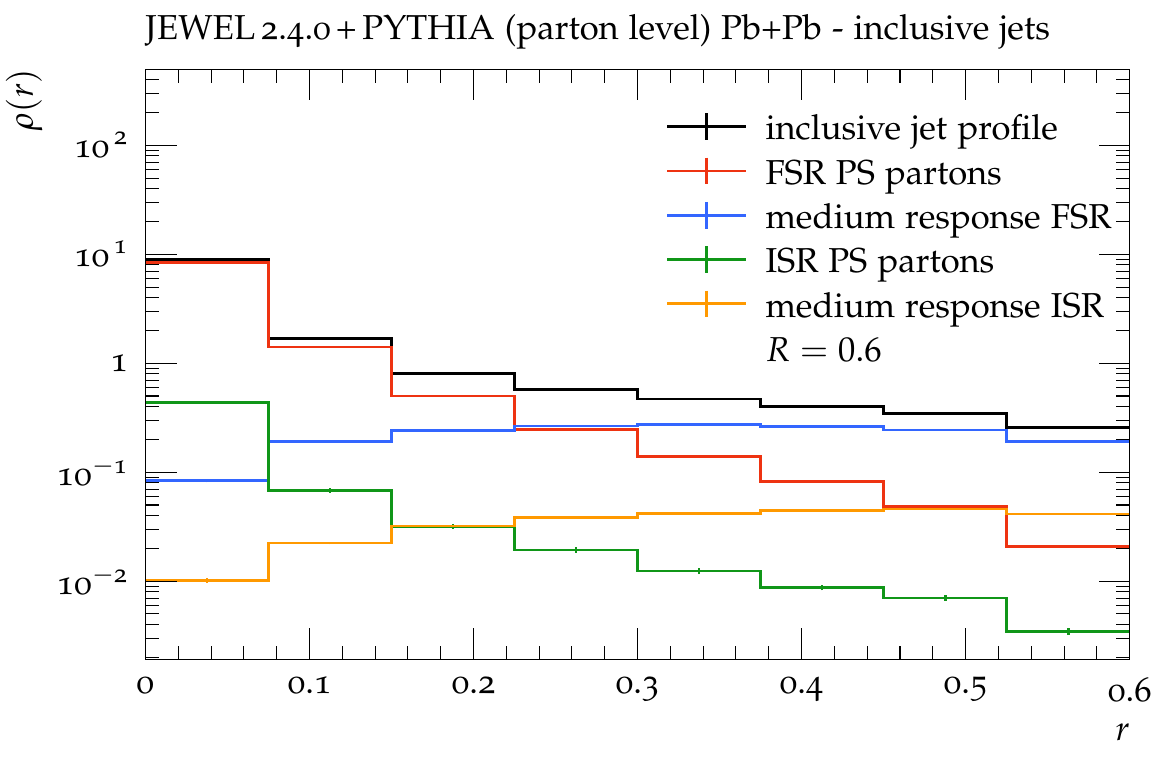}
	\caption{Jet profile for inclusive $R=0.6$ jets in central Pb+Pb collisions and the contributions from partons in the parton showers of initial and final state emissions as well as their respective medium response components.}
	\label{fig:jetprcontribs_PbPb}
\end{figure}

The picture is similar in Pb+Pb collisions (fig.~\ref{fig:jetprofile_PbPb}), but here the jet profile flattens out at large $r$ compared to p+p. In \jewel\ this is entirely due to medium response~\cite{KunnawalkamElayavalli:2017hxo}. 
When plotting the ratio of the ISR contribution to the full jet profile for inclusive jets (figs.~\ref{fig:isrjetprofile_pp} and \ref{fig:isrjetprofile_PbPb}) one sees how the IS component gets more and more important as $R$ and with it the available area increases. In p+p collisions the ISR contribution exceeds $\unit[10]{\%}$ for $R=1$ jets at large $r$. In Pb+Pb collisions it is generally larger, particularly at large $r$, and goes up to $\unit[25]{\%}$. This difference between Pb+Pb and p+p comes mainly from medium response to interaction of ISR partons. To illustrate this point, in fig.~\ref{fig:jetprcontribs_PbPb} the jet profile of inclusive $R=0.6$ jets in Pb+Pb collisions is broken down into the contributions from partons from the parton showers of initial and final state emissions and the respective medium response components. The same pattern is seen in IS and FS contributions: the parton shower partons dominate at small $r$, but at large distance from the jet axis the medium response components start to dominate.

\medskip

The jet profile characterises the momentum distribution around the jet axis in a fairly generic way. A more detailed look into the sub-structure of jets is offered by SoftDrop tagging~\cite{Dasgupta:2013ihk,Larkoski:2014wba}. This procedure, that can also be used to groom away soft contributions, identifies a hard two-prong structure inside the jet. First, the anti-$k_\perp$ jet is re-clustered with the Cambridge/Aachen algorithm~\cite{Wobisch:1998wt}. Then, in an iterative procedure, the clustering is undone splitting the jet into two sub-jets. In each step the softer of the two sub-jets is dropped until a configuration is reached that satisfies
\begin{equation}
	z_g = \frac{\text{min}(p_{\perp,1}, p_{\perp,2})}{p_{\perp,1}+ p_{\perp,2}} > z_{\mathrm{cut}} \left(\frac{\Delta R_{12}}{R}\right)^{\beta} \,,
\end{equation}
where $\Delta R_{12}$ is the angular separation between the two sub-jets and $p_{\perp,i}$ are their transverse momenta. The case $\beta=0$ has received special attention because then the $z_g$-distribution is sensitive to the QCD splitting functions~\cite{Larkoski:2015lea}. For this analysis $\beta=0$ and $z_{\mathrm{cut}}=0.1$ is chosen.

\begin{figure}
	\includegraphics[width=8.5cm]{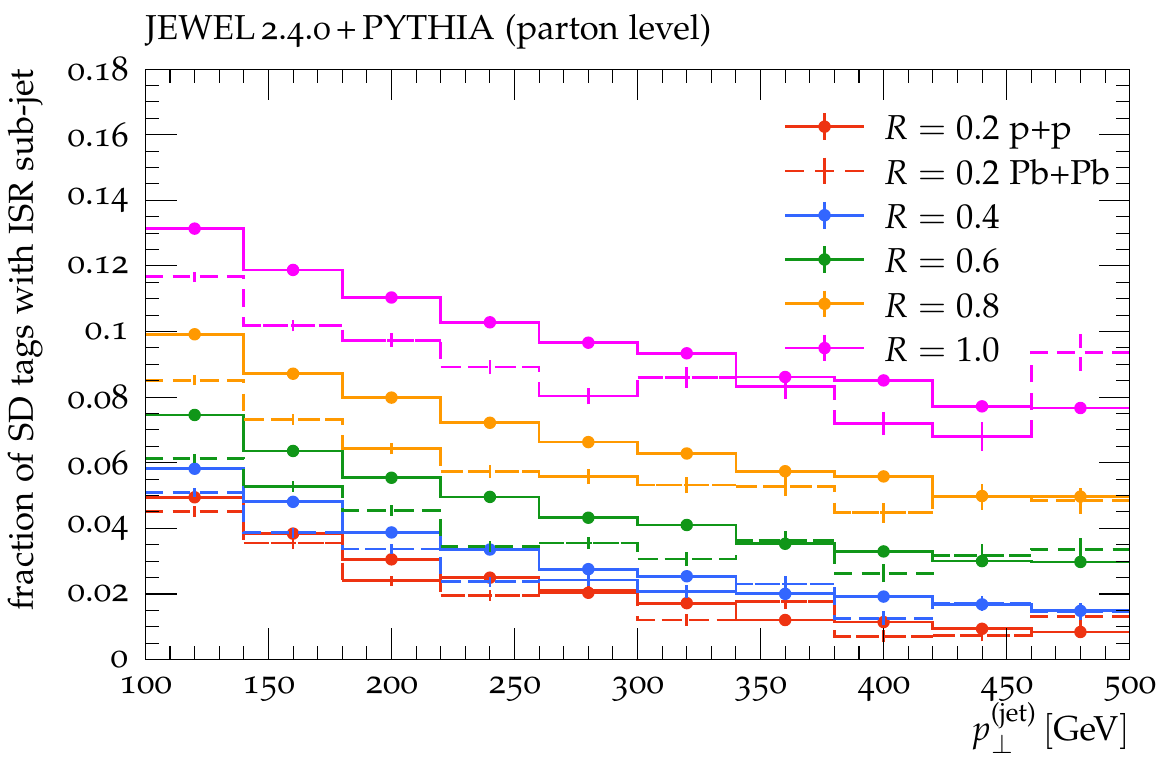}
	\caption{Fraction of SoftDrop ($\beta=0$, $z_{\mathrm{cut}}=0.1$) tagged jets where at least one sub-jet is classified as an IS sub-jet for different jet radii in p+p (solid histograms with markers) and central Pb+Pb collisions (dashed histograms).}
	\label{fig:isfrac_sd}
\end{figure}

Analogously to full jets a sub-jet identified by the SoftDrop algorithm is classified as IS sub-jet if more than half of its $\pt$ is carried by ISR partons. Fig.~\ref{fig:isfrac_sd} shows the fraction of SoftDrop tagged jets that have at least one IS sub-jet (``ISR tagged jets'') as a function of the jet $\pt$ for different jet radii. Again, the fraction decreases with $\jetpt$ and increases with $R$ and is in the range of up to $\unit[10]{\%}$. For small $R$ the majority of ISR tagged jets has two IS sub-jets, i.e.\ is most likely an IS jet, while at large $R$ mixed configurations with one IS and one FS sub-jet dominate. In contrast to the jet profile, the ISR contribution is smaller in Pb+Pb than in p+p collisions. The reason for this is that SoftDrop isolates a hard structure inside the jet, while the jet profile includes contributions from soft large-angle fragments. The response the quenching is thus very different.

\begin{figure}
	\includegraphics[width=8.5cm]{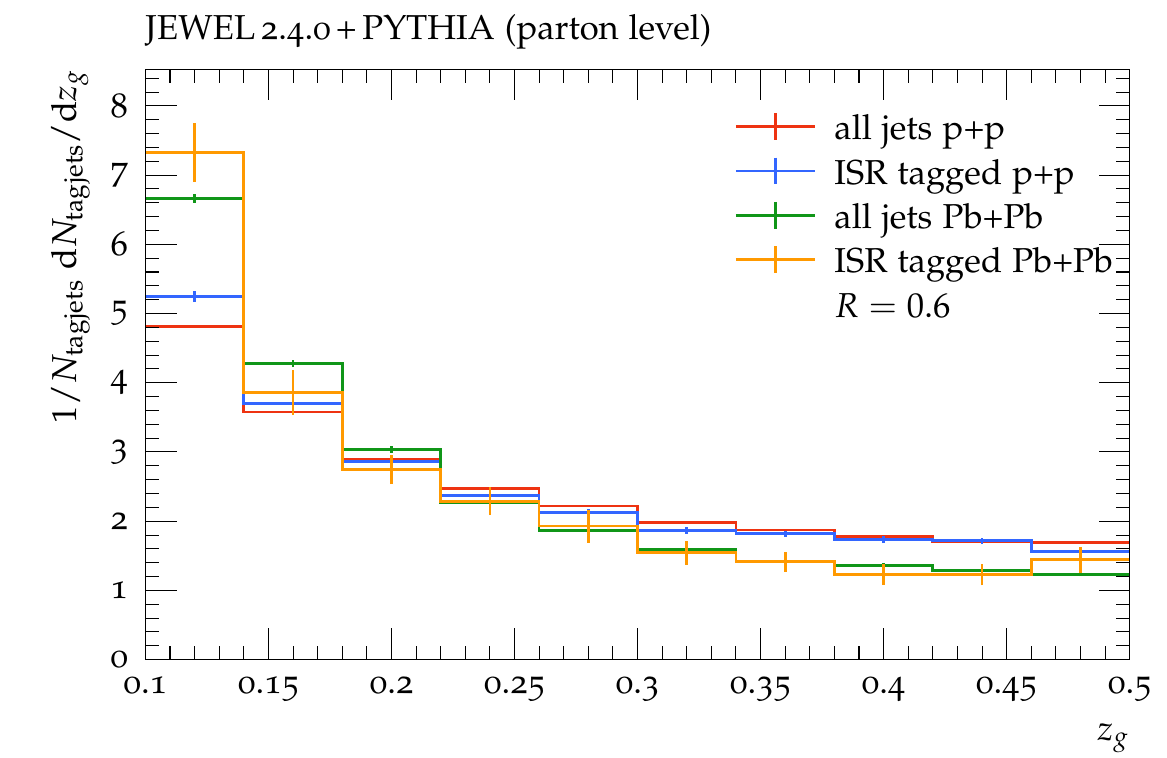}
	\caption{$z_g$-distribution in SoftDrop tagged $R=0.6$ jets for all jets and jets with at least one IS sub-jet (``ISR tagged'') in p+p and Pb+Pb collisions.}
	\label{fig:zg}
\end{figure}

The $z_g$ distribution in ISR tagged jets is very similar to the inclusive distribution (fig.~\ref{fig:zg}) both in p+p and Pb+Pb collisions. This is remarkable, since even in mixed configurations, where one does not expect the two sub-jets to originate from the same splitting, the distribution is very similar. 

\begin{figure}
	\includegraphics[width=8.5cm]{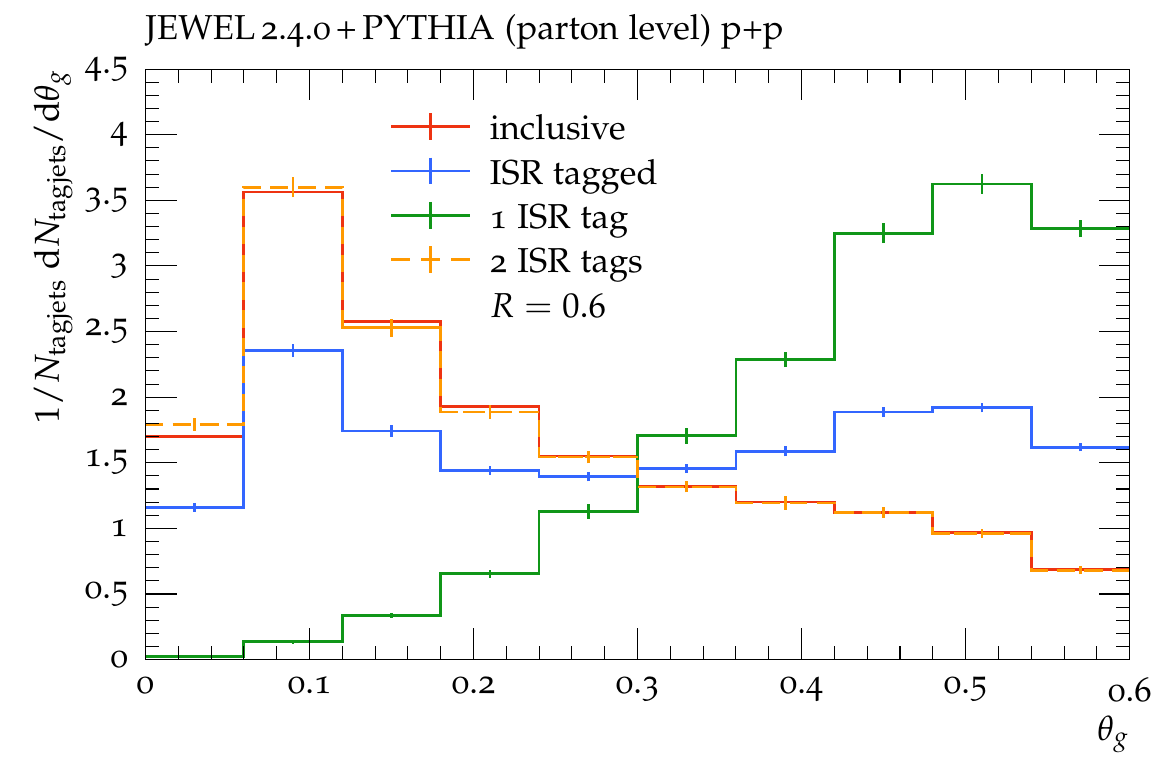}
	\caption{Distribution of opening angle $\theta_g$ between sub-jets in SoftDrop tagged $R=0.6$ jets  in p+p collisions for all jets (red), jets with at least one IS sub-jet (blue), jets with exactly one IS sub-jet (green), and jets with two IS sub-jets (orange).}
	\label{fig:thetag_pp}
\end{figure}

\begin{figure}
	\includegraphics[width=8.5cm]{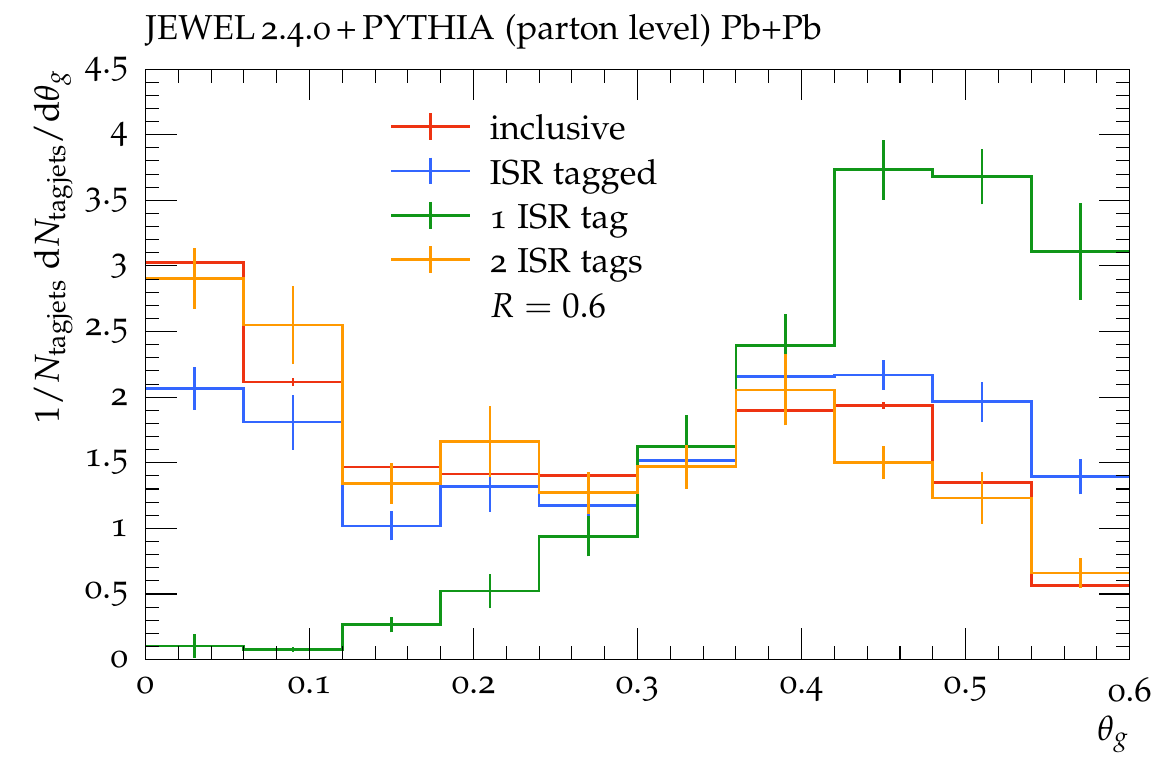}
	\caption{Distribution of opening angle $\theta_g$ between sub-jets in SoftDrop tagged $R=0.6$ jets in central Pb+Pb collisions for all jets (red), jets with at least one IS sub-jet (blue), jets with exactly one IS sub-jet (green), and jets with two IS sub-jets (orange).}
	\label{fig:thetag_PbPb}
\end{figure}

For the opening angle $\theta_g$ between the two sub-jets (figs.~\ref{fig:thetag_pp} and \ref{fig:thetag_PbPb}) the situation is very different: already in p+p collisions the $\theta_g$-distribution develops a bump at large $\theta_g$. This comes from mixed configurations with exactly one IS sub-jet, where it is more likely that the two sub-jets have a large angular separation. This bump looks very similar to the one seen in Pb+Pb collisions in the inclusive distribution, where it is caused by medium response~\cite{Milhano:2017nzm}. This is probably due to the fact that in both cases geometry dictates the probability of finding two un-correlated structures inside a jet. The $\theta_g$-distribution of ISR tagged jets in Pb+Pb does not differ much from the inclusive distribution, only the bump is slightly more pronounced. As in p+p, the distribution of mixed configurations is peaked at large $\theta_g$, while that of double-IS resembles the inclusive distribution. The latter are most likely cases where the two sub-jets actually come from the splitting of a common ancestor. In this case the distribution is dictated by QCD dynamics irrespective of whether the parton at the origin of the branching sequence was radiated from the initial or the final state.

\medskip

\begin{figure}
	\includegraphics[width=8.5cm]{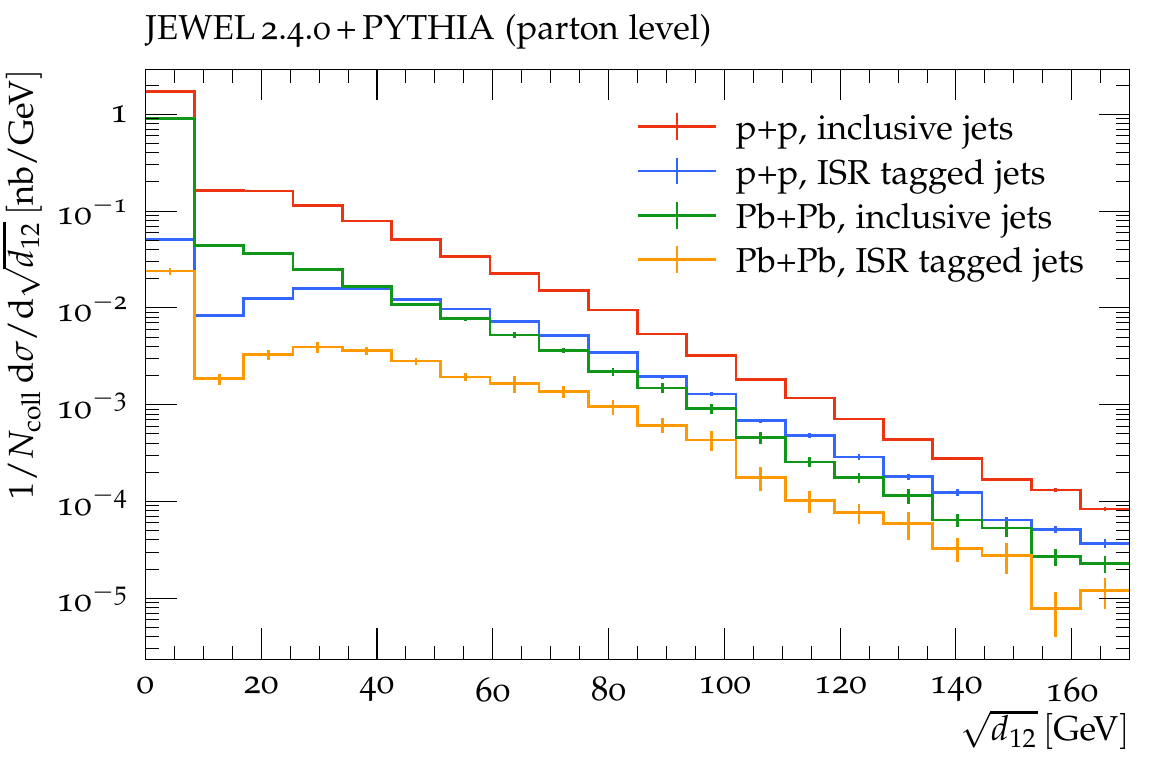}
	\caption{Differential jet cross section per nucleon-nucleon collision for inclusive and ISR tagged jets in p+p and central Pb+Pb collisions.}
	\label{fig:sqrt12}
\end{figure}

The last analysis discussed here does not follow the standard procedure outlined in section~\ref{sec:analysis}, but is motivated by a measurement of large-radius jets found by clustering small-radius jets~\cite{ATLAS:2019rmd}. Following the experimental procedure $R=0.2$ anti-$k_\perp$ jets with $\jetpt > \unit[35]{GeV}$ and $\jeteta < 3$ are found. These are clustered into $R=1$ anti-$k_\perp$ jets with $\jetpt > \unit[158]{GeV}$ and $\jeteta < 2$. The large-radius jets are re-clustered using the $k_\perp$- algorithm~\cite{Ellis:1993tq}. The jet cross section is considered as a function of the distance measure of the last clustering step 
\[ \sqrt{d_{12}} = \min(p_{\perp,1}, p_{\perp,2}) \frac{\Delta R_{12}}{R}
\]
with the aim of finding jets with a hard wide angle splitting. Large-radius jets composed of only a single small-radius jet are defined to have  $\sqrt{d_{12}} = 0$. As before, jets with at least one IS sub-jet after the un-clustering step are labeled as ``ISR tagged jets''.

The $\sqrt{d_{12}}$-distribution is steeply falling with $\sqrt{d_{12}} = 0$ (i.e.\ jets with only one sub-jet) being the most likely value (fig.~\ref{fig:sqrt12}). The fraction of ISR tagged jets first increases with $\sqrt{d_{12}}$ and then levels off between 0.4 and 0.5 with very similar values in p+p and central Pb+Pb collisions. The ISR tagged population is completely dominated by mixed configurations with exactly one IS and one FS sub-jet. The higher values of $\sqrt{d_{12}}$ of these are due to the larger angular separations of the sub-jets than in FS-FS jets. This indicates that in roughly half of the jets with large-$k_\perp$ structures these are due to random combinations.

\begin{figure}
	\includegraphics[width=8.5cm]{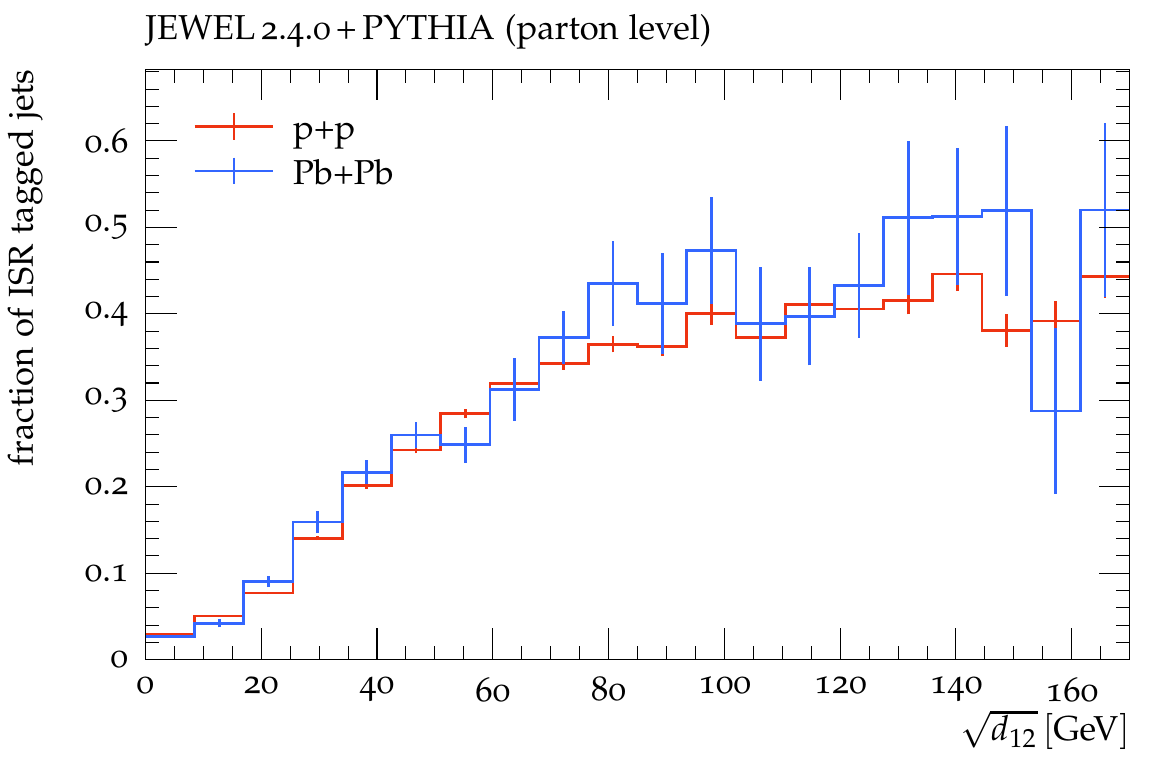}
	\caption{Fraction of ISR tagged jets as a function of the scale $\sqrt{d_{12}}$ of the last $k_\perp$ clustering step.}
	\label{fig:sqrt12_isfrac}
\end{figure}

\section{Conclusions}

Initial state radiation produces jets and it contaminates jets consisting predominantly of final state particles. IS jets are mostly unproblematic in single-inclusive jet samples, because they are ordinary QCD jets. However, they can play a role in observables correlating two or more jets (for example the di-jet acoplanarity shown in fig.~\ref{fig:isfrac_dijets}). 

Sizable ISR contributions to FS jets can be a confounding factor since they share characteristics with medium response: they are weakly correlated with the jet and broadly distributed in angle relative to the jet. This is seen for instance in the jet profile (figs.~\ref{fig:isrjetprofile_pp} and \ref{fig:isrjetprofile_PbPb}) and the $\theta_g$-distribution (figs.~\ref{fig:thetag_pp} and \ref{fig:thetag_PbPb}). 

The exact size of ISR contributions is observable dependent, but the general conclusion from the findings presented in sections~\ref{sec:res1} and \ref{sec:res2} is that they increase with jet radius, as is expected. They are generally small, typically at the level of at most a few percent, in the bulk of the jet population. In the tails of distributions, however, they can become sizable. The jet profile and $\theta_g$-distribution are examples where the ISR contributions can be between $\unit[10]{\%}$ and $\unit[20]{\%}$, i.e.\ at a level where they are relevant when one is aiming for precision modeling. Finally, there are cases like the $\sqrt{d_{12}}$-distribution (fig.~\ref{fig:sqrt12_isfrac}) or the di-jet acoplanarity where the ISR contributions can exceed  $\unit[40]{\%}$. This is not a problem per se, as ISR is ordinary QCD radiation and behaves in much the same way as FSR, but it may require extra effort to include it in the theoretical calculations and models.

\smallskip

Looking at the results presented here from a somewhat different angle, it is worth noting that LO matrix element plus parton shower calculations like \jewel\ are only leading-log accurate in regions of phase space dominated by ISR contributions. For a better theoretical description of these one would have to include matrix element corrections. This technique allows to correct parton shower emissions such that they reproduce exactly (and not only in the collinear limit) the LO multi-leg matrix elements. While this is a standard procedure in p+p physics, it has not been attempted in A+A collisions where the situation is complicated by medium-induced gluon emissions.

\section*{Acknowledgments}

The author would like to thank Guilherme Milhano for carefully reading the manuscript. This study is part of a project that has received funding from the European Research Council (ERC) under the European Union's Horizon 2020 research and innovation programme  (Grant agreement No. 803183, collectiveQCD).

\bibliography{../bib/jetquenching.bib}

\begin{thebibliography}{10}
\expandafter\ifx\csname url\endcsname\relax
  \def\url#1{\texttt{#1}}\fi
\expandafter\ifx\csname urlprefix\endcsname\relax\def\urlprefix{URL }\fi
\expandafter\ifx\csname href\endcsname\relax
  \def\href#1#2{#2} \def\path#1{#1}\fi

\bibitem{Casalderrey-Solana:2007knd}
J.~Casalderrey-Solana, C.~A. Salgado, {Introductory lectures on jet quenching
  in heavy ion collisions}, Acta Phys. Polon. B 38 (2007) 3731--3794.
\newblock \href {http://arxiv.org/abs/0712.3443} {\path{arXiv:0712.3443}}.

\bibitem{Wiedemann:2009sh}
U.~A. Wiedemann, {Jet Quenching in Heavy Ion Collisions} (2009).
\newblock \href {http://arxiv.org/abs/0908.2306} {\path{arXiv:0908.2306}}.

\bibitem{Majumder:2010qh}
A.~Majumder, M.~Van~Leeuwen, {The Theory and Phenomenology of Perturbative QCD
  Based Jet Quenching}, Prog. Part. Nucl. Phys. 66 (2011) 41--92.
\newblock \href {http://arxiv.org/abs/1002.2206} {\path{arXiv:1002.2206}},
  \href {https://doi.org/10.1016/j.ppnp.2010.09.001}
  {\path{doi:10.1016/j.ppnp.2010.09.001}}.

\bibitem{Mehtar-Tani:2013pia}
Y.~Mehtar-Tani, J.~G. Milhano, K.~Tywoniuk, {Jet physics in heavy-ion
  collisions}, Int. J. Mod. Phys. A 28 (2013) 1340013.
\newblock \href {http://arxiv.org/abs/1302.2579} {\path{arXiv:1302.2579}},
  \href {https://doi.org/10.1142/S0217751X13400137}
  {\path{doi:10.1142/S0217751X13400137}}.

\bibitem{Connors:2017ptx}
M.~Connors, C.~Nattrass, R.~Reed, S.~Salur, {Jet measurements in heavy ion
  physics}, Rev. Mod. Phys. 90 (2018) 025005.
\newblock \href {http://arxiv.org/abs/1705.01974} {\path{arXiv:1705.01974}},
  \href {https://doi.org/10.1103/RevModPhys.90.025005}
  {\path{doi:10.1103/RevModPhys.90.025005}}.

\bibitem{Qin:2015srf}
G.-Y. Qin, X.-N. Wang, {Jet quenching in high-energy heavy-ion collisions},
  Int. J. Mod. Phys. E 24~(11) (2015) 1530014.
\newblock \href {http://arxiv.org/abs/1511.00790} {\path{arXiv:1511.00790}},
  \href {https://doi.org/10.1142/S0218301315300143}
  {\path{doi:10.1142/S0218301315300143}}.

\bibitem{Cao:2020wlm}
S.~Cao, X.-N. Wang, {Jet quenching and medium response in high-energy heavy-ion
  collisions: a review}, Rept. Prog. Phys. 84~(2) (2021) 024301.
\newblock \href {http://arxiv.org/abs/2002.04028} {\path{arXiv:2002.04028}},
  \href {https://doi.org/10.1088/1361-6633/abc22b}
  {\path{doi:10.1088/1361-6633/abc22b}}.

\bibitem{Cunqueiro:2021wls}
L.~Cunqueiro, A.~M. Sickles, {Studying the QGP with Jets at the LHC and RHIC},
  Prog. Part. Nucl. Phys. 124 (2022) 103940.
\newblock \href {http://arxiv.org/abs/2110.14490} {\path{arXiv:2110.14490}},
  \href {https://doi.org/10.1016/j.ppnp.2022.103940}
  {\path{doi:10.1016/j.ppnp.2022.103940}}.

\bibitem{Apolinario:2022vzg}
L.~Apolin\'ario, Y.-J. Lee, M.~Winn, {Heavy quarks and jets as probes of the
  QGP}, Prog. Part. Nucl. Phys. 127 (2022) 103990.
\newblock \href {http://arxiv.org/abs/2203.16352} {\path{arXiv:2203.16352}},
  \href {https://doi.org/10.1016/j.ppnp.2022.103990}
  {\path{doi:10.1016/j.ppnp.2022.103990}}.

\bibitem{Casalderrey-Solana:2011ule}
J.~Casalderrey-Solana, E.~Iancu, {Interference effects in medium-induced gluon
  radiation}, JHEP 08 (2011) 015.
\newblock \href {http://arxiv.org/abs/1105.1760} {\path{arXiv:1105.1760}},
  \href {https://doi.org/10.1007/JHEP08(2011)015}
  {\path{doi:10.1007/JHEP08(2011)015}}.

\bibitem{Mehtar-Tani:2011hma}
Y.~Mehtar-Tani, C.~A. Salgado, K.~Tywoniuk, {Jets in QCD Media: From Color
  Coherence to Decoherence}, Phys. Lett. B 707 (2012) 156--159.
\newblock \href {http://arxiv.org/abs/1102.4317} {\path{arXiv:1102.4317}},
  \href {https://doi.org/10.1016/j.physletb.2011.12.042}
  {\path{doi:10.1016/j.physletb.2011.12.042}}.

\bibitem{Blaizot:2014ula}
J.-P. Blaizot, Y.~Mehtar-Tani, M.~A.~C. Torres, {Angular structure of the
  in-medium QCD cascade}, Phys. Rev. Lett. 114~(22) (2015) 222002.
\newblock \href {http://arxiv.org/abs/1407.0326} {\path{arXiv:1407.0326}},
  \href {https://doi.org/10.1103/PhysRevLett.114.222002}
  {\path{doi:10.1103/PhysRevLett.114.222002}}.

\bibitem{Chien:2015hda}
Y.-T. Chien, I.~Vitev, {Towards the understanding of jet shapes and cross
  sections in heavy ion collisions using soft-collinear effective theory}, JHEP
  05 (2016) 023.
\newblock \href {http://arxiv.org/abs/1509.07257} {\path{arXiv:1509.07257}},
  \href {https://doi.org/10.1007/JHEP05(2016)023}
  {\path{doi:10.1007/JHEP05(2016)023}}.

\bibitem{Casalderrey-Solana:2015bww}
J.~Casalderrey-Solana, D.~Pablos, K.~Tywoniuk, {Two-gluon emission and
  interference in a thin QCD medium: insights into jet formation}, JHEP 11
  (2016) 174.
\newblock \href {http://arxiv.org/abs/1512.07561} {\path{arXiv:1512.07561}},
  \href {https://doi.org/10.1007/JHEP11(2016)174}
  {\path{doi:10.1007/JHEP11(2016)174}}.

\bibitem{Dominguez:2019ges}
F.~Dom\'\i{}nguez, J.~G. Milhano, C.~A. Salgado, K.~Tywoniuk, V.~Vila, {Mapping
  collinear in-medium parton splittings}, Eur. Phys. J. C 80~(1) (2020) 11.
\newblock \href {http://arxiv.org/abs/1907.03653} {\path{arXiv:1907.03653}},
  \href {https://doi.org/10.1140/epjc/s10052-019-7563-0}
  {\path{doi:10.1140/epjc/s10052-019-7563-0}}.

\bibitem{Arnold:2020uzm}
P.~Arnold, T.~Gorda, S.~Iqbal, {The LPM effect in sequential bremsstrahlung:
  nearly complete results for QCD}, JHEP 11 (2020) 053.
\newblock \href {http://arxiv.org/abs/2007.15018} {\path{arXiv:2007.15018}},
  \href {https://doi.org/10.1007/JHEP11(2020)053}
  {\path{doi:10.1007/JHEP11(2020)053}}.

\bibitem{Mehtar-Tani:2021fud}
Y.~Mehtar-Tani, D.~Pablos, K.~Tywoniuk, {Cone-Size Dependence of Jet
  Suppression in Heavy-Ion Collisions}, Phys. Rev. Lett. 127~(25) (2021)
  252301.
\newblock \href {http://arxiv.org/abs/2101.01742} {\path{arXiv:2101.01742}},
  \href {https://doi.org/10.1103/PhysRevLett.127.252301}
  {\path{doi:10.1103/PhysRevLett.127.252301}}.

\bibitem{Barata:2021byj}
J.~a. Barata, F.~Dom\'\i{}nguez, C.~A. Salgado, V.~Vila, {A modified in-medium
  evolution equation with color coherence}, JHEP 05 (2021) 148.
\newblock \href {http://arxiv.org/abs/2101.12135} {\path{arXiv:2101.12135}},
  \href {https://doi.org/10.1007/JHEP05(2021)148}
  {\path{doi:10.1007/JHEP05(2021)148}}.

\bibitem{Andres:2022ndd}
C.~Andres, F.~Dominguez, A.~V. Sadofyev, C.~A. Salgado, {Jet Broadening in
  Flowing Matter -- Resummation} (7 2022).
\newblock \href {http://arxiv.org/abs/2207.07141} {\path{arXiv:2207.07141}}.

\bibitem{Neufeld:2011yh}
R.~B. Neufeld, I.~Vitev, {Parton showers as sources of energy-momentum
  deposition in the QGP and their implication for shockwave formation at RHIC
  and at the LHC}, Phys. Rev. C 86 (2012) 024905.
\newblock \href {http://arxiv.org/abs/1105.2067} {\path{arXiv:1105.2067}},
  \href {https://doi.org/10.1103/PhysRevC.86.024905}
  {\path{doi:10.1103/PhysRevC.86.024905}}.

\bibitem{Tachibana:2014lja}
Y.~Tachibana, T.~Hirano, {Momentum transport away from a jet in an expanding
  nuclear medium}, Phys. Rev. C 90~(2) (2014) 021902.
\newblock \href {http://arxiv.org/abs/1402.6469} {\path{arXiv:1402.6469}},
  \href {https://doi.org/10.1103/PhysRevC.90.021902}
  {\path{doi:10.1103/PhysRevC.90.021902}}.

\bibitem{He:2015pra}
Y.~He, T.~Luo, X.-N. Wang, Y.~Zhu, {Linear Boltzmann Transport for Jet
  Propagation in the Quark-Gluon Plasma: Elastic Processes and Medium Recoil},
  Phys. Rev. C91 (2015) 054908.
\newblock \href {http://arxiv.org/abs/1503.03313} {\path{arXiv:1503.03313}},
  \href {https://doi.org/10.1103/PhysRevC.91.054908}
  {\path{doi:10.1103/PhysRevC.91.054908}}.

\bibitem{Wang:2013cia}
X.-N. Wang, Y.~Zhu, {Medium Modification of $\gamma$-jets in High-energy
  Heavy-ion Collisions}, Phys. Rev. Lett. 111~(6) (2013) 062301.
\newblock \href {http://arxiv.org/abs/1302.5874} {\path{arXiv:1302.5874}},
  \href {https://doi.org/10.1103/PhysRevLett.111.062301}
  {\path{doi:10.1103/PhysRevLett.111.062301}}.

\bibitem{Cao:2016gvr}
S.~Cao, T.~Luo, G.-Y. Qin, X.-N. Wang, {Linearized Boltzmann transport model
  for jet propagation in the quark-gluon plasma: Heavy quark evolution}, Phys.
  Rev. C 94~(1) (2016) 014909.
\newblock \href {http://arxiv.org/abs/1605.06447} {\path{arXiv:1605.06447}},
  \href {https://doi.org/10.1103/PhysRevC.94.014909}
  {\path{doi:10.1103/PhysRevC.94.014909}}.

\bibitem{Casalderrey-Solana:2016jvj}
J.~Casalderrey-Solana, D.~Gulhan, G.~Milhano, D.~Pablos, K.~Rajagopal, {Angular
  Structure of Jet Quenching Within a Hybrid Strong/Weak Coupling Model}, JHEP
  03 (2017) 135.
\newblock \href {http://arxiv.org/abs/1609.05842} {\path{arXiv:1609.05842}},
  \href {https://doi.org/10.1007/JHEP03(2017)135}
  {\path{doi:10.1007/JHEP03(2017)135}}.

\bibitem{KunnawalkamElayavalli:2017hxo}
R.~Kunnawalkam~Elayavalli, K.~C. Zapp, {Medium response in JEWEL and its impact
  on jet shape observables in heavy ion collisions}, JHEP 07 (2017) 141.
\newblock \href {http://arxiv.org/abs/1707.01539} {\path{arXiv:1707.01539}},
  \href {https://doi.org/10.1007/JHEP07(2017)141}
  {\path{doi:10.1007/JHEP07(2017)141}}.

\bibitem{hilmisthesis}
H.~Kycyku, The impact of initial state radiation on jets from pbpb collisions,
  student Paper (2021).

\bibitem{CMS:2021vui}
A.~M. Sirunyan, et~al., {First measurement of large area jet transverse
  momentum spectra in heavy-ion collisions}, JHEP 05 (2021) 284.
\newblock \href {http://arxiv.org/abs/2102.13080} {\path{arXiv:2102.13080}},
  \href {https://doi.org/10.1007/JHEP05(2021)284}
  {\path{doi:10.1007/JHEP05(2021)284}}.

\bibitem{ATLAS:2019rmd}
{Measurement of suppression of large-radius jets and its dependence on
  substructure in Pb+Pb at 5.02 TeV by ATLAS detector} (11 2019).

\bibitem{Armesto:2008qh}
N.~Armesto, L.~Cunqueiro, C.~A. Salgado, {Implementation of a medium-modified
  parton shower algorithm}, Eur. Phys. J. C 61 (2009) 775--778.
\newblock \href {http://arxiv.org/abs/0809.4433} {\path{arXiv:0809.4433}},
  \href {https://doi.org/10.1140/epjc/s10052-008-0824-y}
  {\path{doi:10.1140/epjc/s10052-008-0824-y}}.

\bibitem{Schenke:2009gb}
B.~Schenke, C.~Gale, S.~Jeon, {MARTINI: An event generator for relativistic
  heavy-ion collisions}, Phys. Rev. C80 (2009) 054913.
\newblock \href {http://arxiv.org/abs/0909.2037} {\path{arXiv:0909.2037}},
  \href {https://doi.org/10.1103/PhysRevC.80.054913}
  {\path{doi:10.1103/PhysRevC.80.054913}}.

\bibitem{Casalderrey-Solana:2014bpa}
J.~Casalderrey-Solana, D.~C. Gulhan, J.~G. Milhano, D.~Pablos, K.~Rajagopal, {A
  Hybrid Strong/Weak Coupling Approach to Jet Quenching}, JHEP 10 (2014) 019,
  [Erratum: JHEP 09, 175 (2015)].
\newblock \href {http://arxiv.org/abs/1405.3864} {\path{arXiv:1405.3864}},
  \href {https://doi.org/10.1007/JHEP09(2015)175}
  {\path{doi:10.1007/JHEP09(2015)175}}.

\bibitem{Chen:2017zte}
W.~Chen, S.~Cao, T.~Luo, L.-G. Pang, X.-N. Wang, {Effects of jet-induced medium
  excitation in $\gamma$-hadron correlation in A+A collisions}, Phys. Lett. B
  777 (2018) 86--90.
\newblock \href {http://arxiv.org/abs/1704.03648} {\path{arXiv:1704.03648}},
  \href {https://doi.org/10.1016/j.physletb.2017.12.015}
  {\path{doi:10.1016/j.physletb.2017.12.015}}.

\bibitem{Putschke:2019yrg}
J.~H. Putschke, et~al., {The JETSCAPE framework} (3 2019).
\newblock \href {http://arxiv.org/abs/1903.07706} {\path{arXiv:1903.07706}}.

\bibitem{Ke:2020clc}
W.~Ke, X.-N. Wang, {QGP modification to single inclusive jets in a calibrated
  transport model}, JHEP 05 (2021) 041.
\newblock \href {http://arxiv.org/abs/2010.13680} {\path{arXiv:2010.13680}},
  \href {https://doi.org/10.1007/JHEP05(2021)041}
  {\path{doi:10.1007/JHEP05(2021)041}}.

\bibitem{Zapp:2012ak}
K.~C. Zapp, F.~Krauss, U.~A. Wiedemann, {A perturbative framework for jet
  quenching}, JHEP 1303 (2013) 080.
\newblock \href {http://arxiv.org/abs/1212.1599} {\path{arXiv:1212.1599}},
  \href {https://doi.org/10.1007/JHEP03(2013)080}
  {\path{doi:10.1007/JHEP03(2013)080}}.

\bibitem{Zapp:2013vla}
K.~C. Zapp, {JEWEL 2.0.0: directions for use}, Eur.Phys.J. C74 (2014) 2762.
\newblock \href {http://arxiv.org/abs/1311.0048} {\path{arXiv:1311.0048}},
  \href {https://doi.org/10.1140/epjc/s10052-014-2762-1}
  {\path{doi:10.1140/epjc/s10052-014-2762-1}}.

\bibitem{Sjostrand:2006za}
T.~Sjostrand, S.~Mrenna, P.~Skands, {PYTHIA 6.4 physics and manual}, JHEP 05
  (2006) 026.
\newblock \href {http://arxiv.org/abs/hep-ph/0603175}
  {\path{arXiv:hep-ph/0603175}}.

\bibitem{Eskola:2016oht}
K.~J. Eskola, P.~Paakkinen, H.~Paukkunen, C.~A. Salgado, {EPPS16: Nuclear
  parton distributions with LHC data}, Eur. Phys. J. C 77~(3) (2017) 163.
\newblock \href {http://arxiv.org/abs/1612.05741} {\path{arXiv:1612.05741}},
  \href {https://doi.org/10.1140/epjc/s10052-017-4725-9}
  {\path{doi:10.1140/epjc/s10052-017-4725-9}}.

\bibitem{Dulat:2015mca}
S.~Dulat, T.-J. Hou, J.~Gao, M.~Guzzi, J.~Huston, P.~Nadolsky, J.~Pumplin,
  C.~Schmidt, D.~Stump, C.~P. Yuan, {New parton distribution functions from a
  global analysis of quantum chromodynamics}, Phys. Rev. D 93~(3) (2016)
  033006.
\newblock \href {http://arxiv.org/abs/1506.07443} {\path{arXiv:1506.07443}},
  \href {https://doi.org/10.1103/PhysRevD.93.033006}
  {\path{doi:10.1103/PhysRevD.93.033006}}.

\bibitem{Buckley:2014ana}
A.~Buckley, J.~Ferrando, S.~Lloyd, K.~Nordstr\"om, B.~Page, M.~R\"ufenacht,
  M.~Sch\"onherr, G.~Watt, {LHAPDF6: parton density access in the LHC precision
  era}, Eur. Phys. J. C 75 (2015) 132.
\newblock \href {http://arxiv.org/abs/1412.7420} {\path{arXiv:1412.7420}},
  \href {https://doi.org/10.1140/epjc/s10052-015-3318-8}
  {\path{doi:10.1140/epjc/s10052-015-3318-8}}.

\bibitem{Milhano:2022kzx}
J.~G. Milhano, K.~Zapp, {Improved background subtraction and a fresh look at
  jet sub-structure in JEWEL} (7 2022).
\newblock \href {http://arxiv.org/abs/2207.14814} {\path{arXiv:2207.14814}}.

\bibitem{Shen:2014vra}
C.~Shen, Z.~Qiu, H.~Song, J.~Bernhard, S.~Bass, U.~Heinz, {The iEBE-VISHNU code
  package for relativistic heavy-ion collisions}, Comput. Phys. Commun. 199
  (2016) 61--85.
\newblock \href {http://arxiv.org/abs/1409.8164} {\path{arXiv:1409.8164}},
  \href {https://doi.org/10.1016/j.cpc.2015.08.039}
  {\path{doi:10.1016/j.cpc.2015.08.039}}.

\bibitem{Bierlich:2019rhm}
C.~Bierlich, et~al., {Robust Independent Validation of Experiment and Theory:
  Rivet version 3}, SciPost Phys. 8 (2020) 026.
\newblock \href {http://arxiv.org/abs/1912.05451} {\path{arXiv:1912.05451}},
  \href {https://doi.org/10.21468/SciPostPhys.8.2.026}
  {\path{doi:10.21468/SciPostPhys.8.2.026}}.

\bibitem{Cacciari:2008gp}
M.~Cacciari, G.~P. Salam, G.~Soyez, {The anti-$k_t$ jet clustering algorithm},
  JHEP 04 (2008) 063.
\newblock \href {http://arxiv.org/abs/0802.1189} {\path{arXiv:0802.1189}},
  \href {https://doi.org/10.1088/1126-6708/2008/04/063}
  {\path{doi:10.1088/1126-6708/2008/04/063}}.

\bibitem{Cacciari:2011ma}
M.~Cacciari, G.~P. Salam, G.~Soyez, {FastJet User Manual}, Eur. Phys. J. C 72
  (2012) 1896.
\newblock \href {http://arxiv.org/abs/1111.6097} {\path{arXiv:1111.6097}},
  \href {https://doi.org/10.1140/epjc/s10052-012-1896-2}
  {\path{doi:10.1140/epjc/s10052-012-1896-2}}.

\bibitem{DEramo:2018eoy}
F.~D'Eramo, K.~Rajagopal, Y.~Yin, {Moli\`ere scattering in quark-gluon plasma:
  finding point-like scatterers in a liquid}, JHEP 01 (2019) 172.
\newblock \href {http://arxiv.org/abs/1808.03250} {\path{arXiv:1808.03250}},
  \href {https://doi.org/10.1007/JHEP01(2019)172}
  {\path{doi:10.1007/JHEP01(2019)172}}.

\bibitem{Dasgupta:2013ihk}
M.~Dasgupta, A.~Fregoso, S.~Marzani, G.~P. Salam, {Towards an understanding of
  jet substructure}, JHEP 09 (2013) 029.
\newblock \href {http://arxiv.org/abs/1307.0007} {\path{arXiv:1307.0007}},
  \href {https://doi.org/10.1007/JHEP09(2013)029}
  {\path{doi:10.1007/JHEP09(2013)029}}.

\bibitem{Larkoski:2014wba}
A.~J. Larkoski, S.~Marzani, G.~Soyez, J.~Thaler, {Soft Drop}, JHEP 05 (2014)
  146.
\newblock \href {http://arxiv.org/abs/1402.2657} {\path{arXiv:1402.2657}},
  \href {https://doi.org/10.1007/JHEP05(2014)146}
  {\path{doi:10.1007/JHEP05(2014)146}}.

\bibitem{Wobisch:1998wt}
M.~Wobisch, T.~Wengler, {Hadronization corrections to jet cross-sections in
  deep inelastic scattering}, in: {Workshop on Monte Carlo Generators for HERA
  Physics (Plenary Starting Meeting)}, 1998, pp. 270--279.
\newblock \href {http://arxiv.org/abs/hep-ph/9907280}
  {\path{arXiv:hep-ph/9907280}}.

\bibitem{Larkoski:2015lea}
A.~J. Larkoski, S.~Marzani, J.~Thaler, {Sudakov Safety in Perturbative QCD},
  Phys. Rev. D91~(11) (2015) 111501.
\newblock \href {http://arxiv.org/abs/1502.01719} {\path{arXiv:1502.01719}},
  \href {https://doi.org/10.1103/PhysRevD.91.111501}
  {\path{doi:10.1103/PhysRevD.91.111501}}.

\bibitem{Milhano:2017nzm}
G.~Milhano, U.~A. Wiedemann, K.~C. Zapp, {Sensitivity of jet substructure to
  jet-induced medium response}, Phys. Lett. B 779 (2018) 409--413.
\newblock \href {http://arxiv.org/abs/1707.04142} {\path{arXiv:1707.04142}},
  \href {https://doi.org/10.1016/j.physletb.2018.01.029}
  {\path{doi:10.1016/j.physletb.2018.01.029}}.

\bibitem{Ellis:1993tq}
S.~D. Ellis, D.~E. Soper, {Successive combination jet algorithm for hadron
  collisions}, Phys. Rev. D 48 (1993) 3160--3166.
\newblock \href {http://arxiv.org/abs/hep-ph/9305266}
  {\path{arXiv:hep-ph/9305266}}, \href
  {https://doi.org/10.1103/PhysRevD.48.3160}
  {\path{doi:10.1103/PhysRevD.48.3160}}.

\end{thebibliography}

\end{document}